\documentclass[%
 aip,
rsi,%
 amsmath,amssymb,
 reprint,%
]{revtex4-1}
\usepackage{siunitx}
\usepackage{graphicx}% Include figure files
\usepackage{dcolumn}% Align table columns on decimal point
\usepackage{bm}% bold math
\begin{document}

\preprint{APS/123-QED}

\title{Stimulated Brillouin Amplification with Flying Focus}% Force line breaks with \\
%\thanks{Footnote to title of article.}
\author{Zhaohui Wu}
\affiliation{National Key Laboratory of Plasma Physics, Research Center of Laser Fusion, China Academy of Engineering Physics, Mianyang, Sichuan, 621900, China}
\author{Xiaoming Zeng}
\author{Zhaoli Li}
\author{Xiaodong Wang}
\author{Xiao Wang}%
\author{Jie Mu}
\author{Yanlei Zuo}%
\email{zuoyanlei@tsinghua.org.cn}
\author{Kainan Zhou}%
\affiliation{National Key Laboratory of Plasma Physics, Research Center of Laser Fusion, China Academy of Engineering Physics, Mianyang, Sichuan, 621900, China}
\author{Hao Peng}
\email{penghao@sztu.edu.cn}
\affiliation{Shenzhen Key Laboratory of Ultraintense Laser and Advanced Material Technology,
Center for Advanced Material Diagnostic Technology, and College of Engineering Physics,
Shenzhen Technology University,Shenzhen, 518118, China}
\author{C. Riconda}
\affiliation{LULI, Sorbonne Université, CNRS, École Polytechnique, CEA, F-75005, Paris, France}
\author{S. Weber}
\email{Stefan.Weber@eli-beams.eu}
\affiliation{ELI Beamlines facility, Extreme Light Infrastructure ERIC, 25241 Dolni Brezany, Czech Republic}

\date{\today}% It is always \today, today,

\begin{abstract}
 Material damage thresholds pose a fundamental limit to chirped pulse amplification (CPA) in high-power laser systems. Plasma-based amplification via stimulated Brillouin scattering (SBS) offers a damage-free alternative, yet its effectiveness has been hindered by instabilities that constrain interaction length. In this study, we report the first experimental demonstration of SBS amplification driven by a flying focus in a 3-mm plasma channel. The flying focus is generated using chromatic aberration from spherical lenses, with its velocity precisely measured by an interferometric ionization method achieving 6.6 fs timing resolution. At a focus velocity near $-c$, SBS amplification is realized at pump and seed intensities more than two orders of magnitude lower than in conventional setups, yielding a conversion efficiency of 14.5\%. These results validate flying focus as a powerful tool for extending interaction lengths and enabling efficient plasma-based laser amplification at reduced intensities.
\end{abstract}

%Use showkeys class option if keyword
                              %display desired
\maketitle
\section{Introduction}

The advent of chirped pulse amplification (CPA)\cite{Mourou85} represents a milestone in the field of high-power laser technology. Modern CPA-based laser systems have demonstrated the ability to deliver peak powers up to 10 PW ($1~\rm{PW}=10^{15}\rm{W}$), achieving focused intensities that exceed $10^{22}\rm{W/cm^2}$. Nevertheless, further scaling of laser power is fundamentally constrained by the damage threshold of optical materials, particularly at the final compression grating stage. This limitation may be overcome through the implementation of plasma-based amplification schemes, which are intrinsically resistant to damage from intense laser fields. Among these, plasma amplifiers based on stimulated backward Raman scattering (SBRS)\cite{Malkin991,Malkin00,Ping04,Cheng05,Ren08,JUN07,Turnbull12,Wu20} and strongly-coupled stimulated Brillouin scattering (scSBS)\cite{Andreev06,Weber13,Lancia10,Lancia16,Marques19} have been proposed to facilitate the efficient transfer of energy from a long-duration pump pulse to a counter-propagating short seed pulse. Theoretical investigations indicate that such schemes can amplify the unfocused seed to intensities exceeding $10^{17}\rm{W/cm^2}$, suggesting the potential to achieve peak laser powers on the order of $10^{18}\rm{W}$ using apertures of only centimeter scale.
 
Both SBRS and scSBS in plasma have been demonstrated as effective mechanisms for laser amplification\cite{Cheng05,JUN07,Lancia16,Marques19}. However, progress in these approaches has been significantly hindered in recent years. A primary limitation arises from plasma instabilities—particularly thermal effects and the generation of precursors—as the pump pulse traverses the plasma channel prior to interacting with the seed. These instabilities reduce the effective interaction length, with their impact becoming more pronounced in extended plasma channels. For SBRS, saturation effects have been observed as the plasma channel length increases from 2 mm to 4 mm\cite{Turnbull12,Turnbull12s,Wu18}. In the case of scSBS, filamentation instabilities induced by high-intensity laser pulses limit the interaction length to approximately $\sim200~\mu$m\cite{Marques19}. As a result, the available interaction length remains insufficient to simultaneously achieve high energy transfer efficiency and ultrashort compressed pulse durations.

The concept of flying focus has emerged as a promising approach to overcome the aforementioned limitations\cite{Turbunll18,Turbunll18s,Sainte,Dustin19,Turbunll19}. In this scheme, the pump focus propagates with a velocity of $-c$, allowing it to precede the arrival of the seed pulse. This configuration effectively suppresses thermal effects and the formation of precursors, thereby enabling a significant extension of the interaction length and, consequently, improved energy transfer efficiency. Moreover, operating at reduced laser intensities in both SBRS and scSBS regimes provides an additional strategy to alleviate plasma instabilities. These advantages have been theoretically substantiated through numerical simulations\cite{Turbunll18,Turbunll19}. However, experimental confirmation of the efficacy of flying focus in plasma-based amplification remains an open challenge.

Flying focus has previously been realized through the use of diffractive lenses to focus a chirped laser pulse, with the focus velocity controlled via adjustments to the compressor gratings and characterized using a streak camera\cite{Dustin18}. However, the temporal resolution of the streak camera—limited to several picoseconds—has imposed constraints on the precision of such measurements. In the present study, we demonstrate the generation of flying focuses using accumulated chromatic aberration introduced by spherical lenses, with velocity characterization achieved through the technique of interfering ionization. This approach enables the realization of focus velocities ranging from subluminal to superluminal. Crucially, the femtosecond-scale response time of the interfering ionization method results in a substantial improvement in measurement accuracy. With a focus velocity approaching $-c$, the flying focus was applied to stimulated Brillouin scattering (SBS) in a 3 mm plasma channel, where characteristic features of plasma-based SBS were systematically investigated.

\section{Experimental setup}

\begin{figure}[htpb]
\centering
\includegraphics[width=3.5 in]{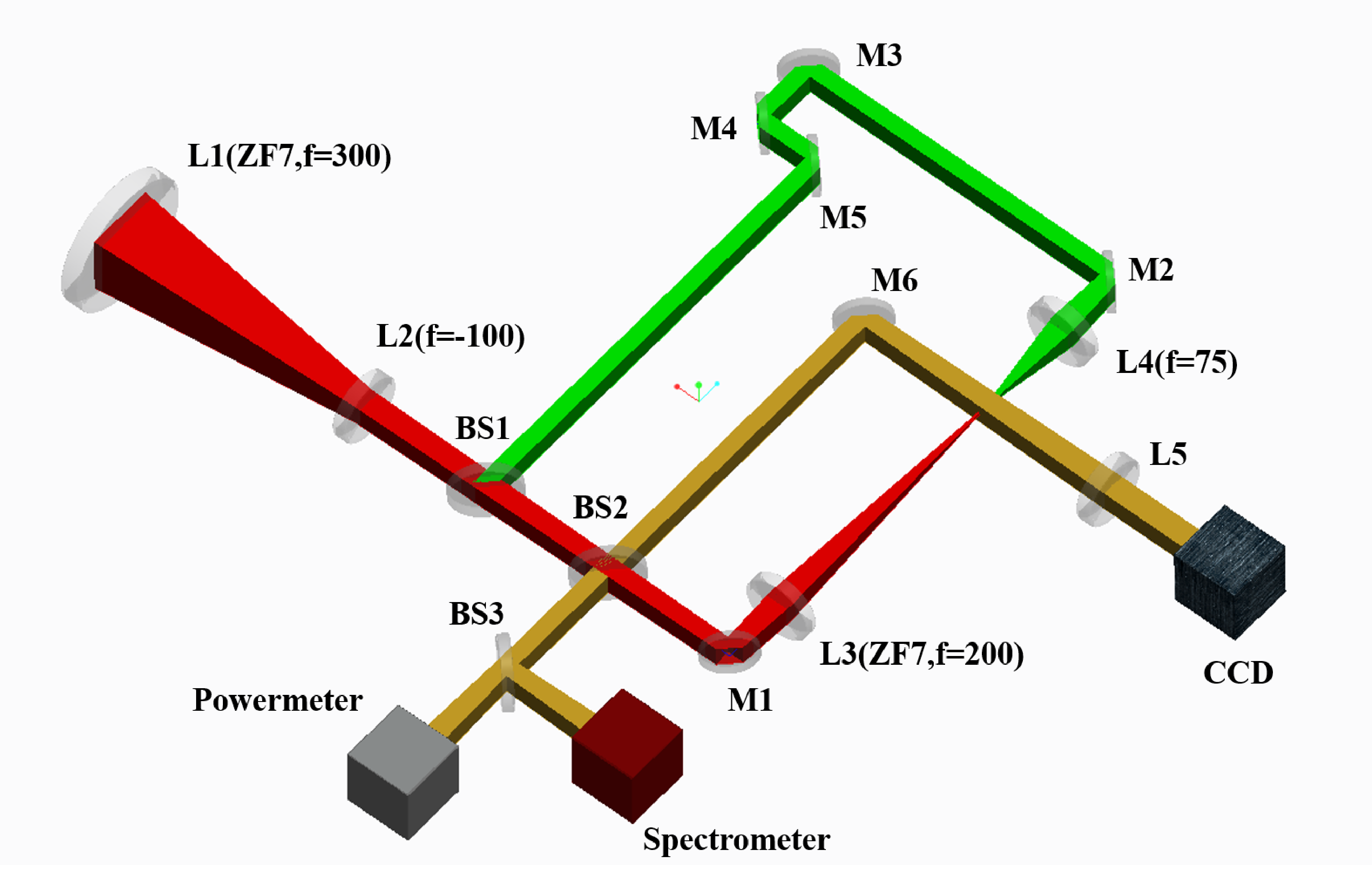}
\caption{Experimental setup for stimulated Brillouin scattering with flying focus.
\label{fig:setup1}}
\end{figure}

The experimental configuration depicted in Fig.\ref{fig:setup1} is outlined as follows: Initially, a laser pulse characterized by a central wavelength of 800 nm and a bandwidth of 80 nm is generated by the Ti:sapphire CPA system. Subsequently, this pulse is condensed from 50 mm to 17 mm through a telescope comprising a ZF7-glass lens with a focal length of $f=300$ mm and a K9-glass lens with $f=-100$ mm. Notably, ZF7 glass is selected for the positive lens due to its large chromatic aberration. The pulse is then divided into two beams by SB1, designated for the pump (red beam) and the probe (green beam) respectively. The pump beam undergoes focusing by a second ZF7-glass lens with $f=200$ mm, resulting in the creation of a line focus with a length of approximately 3 mm. With different pulse chirps, the corresponding focus velocity can be artificially adjusted\cite{Dustin18,Turbunll18s}. Meanwhile, the probe beam is elongated by a delay line and subsequently directed to an achromatic lens with $f=75$ mm, establishing a nearly static focus. This static focus interferes with each point of the pump focus (line focus) to facilitate the measurement of focus velocity. Moreover, the probe beam serves as the seed for SBS when the flying focus approaches a velocity close to $-c$. Additionally, a split mirror (SB2) is incorporated to reflect a portion of the amplified seed towards the power meter and spectrometer for the measurement of pulse energy and spectrum respectively. Furthermore, SB2 splits a minor fraction of the pump beam to serve as the diagnostic pulse, transversely traversing the plasma channel to observe plasma profiles.

\begin{figure}[htpb]
\centering
\includegraphics[width=3 in]{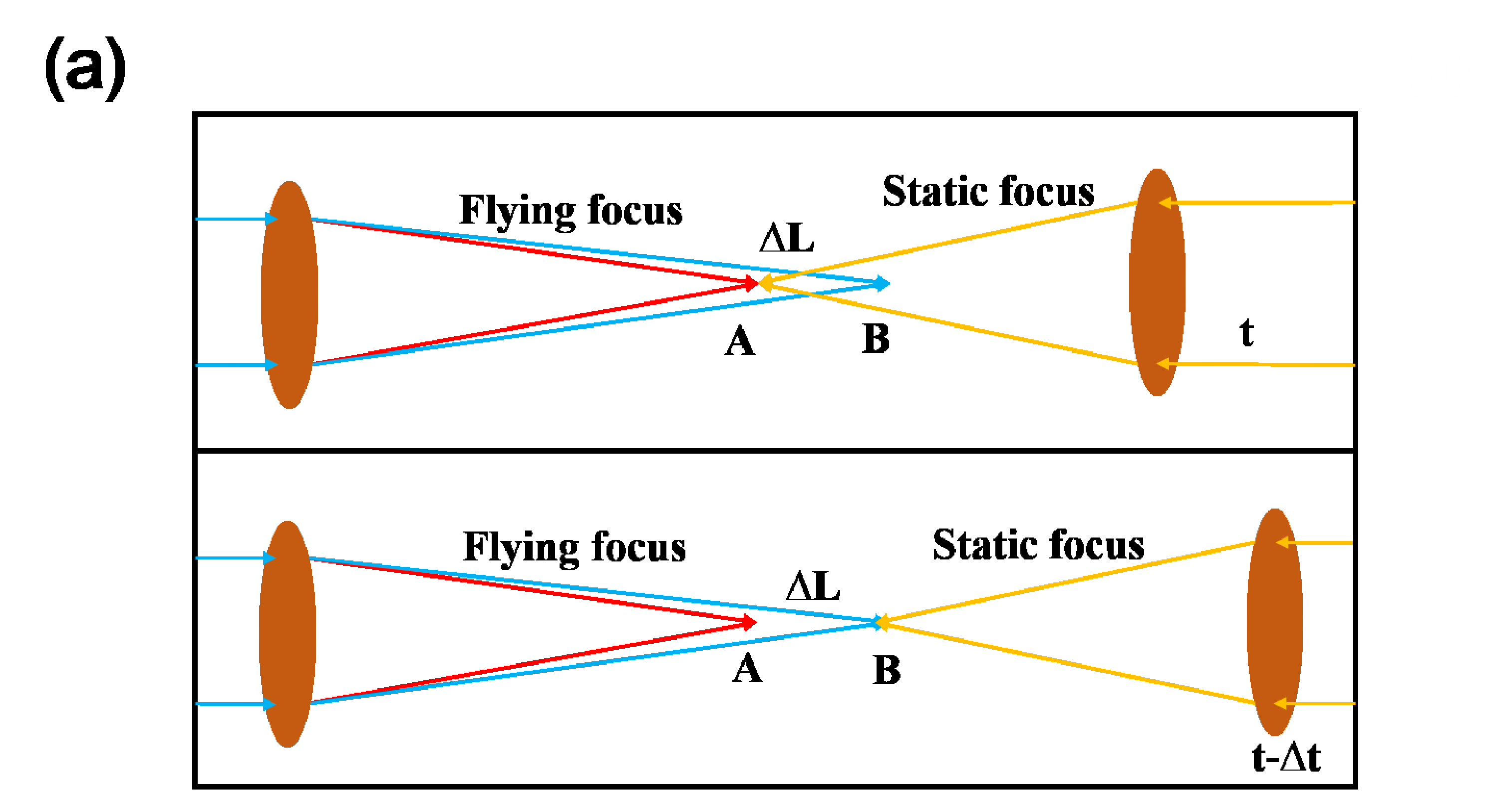}\\
\includegraphics[width=3 in]{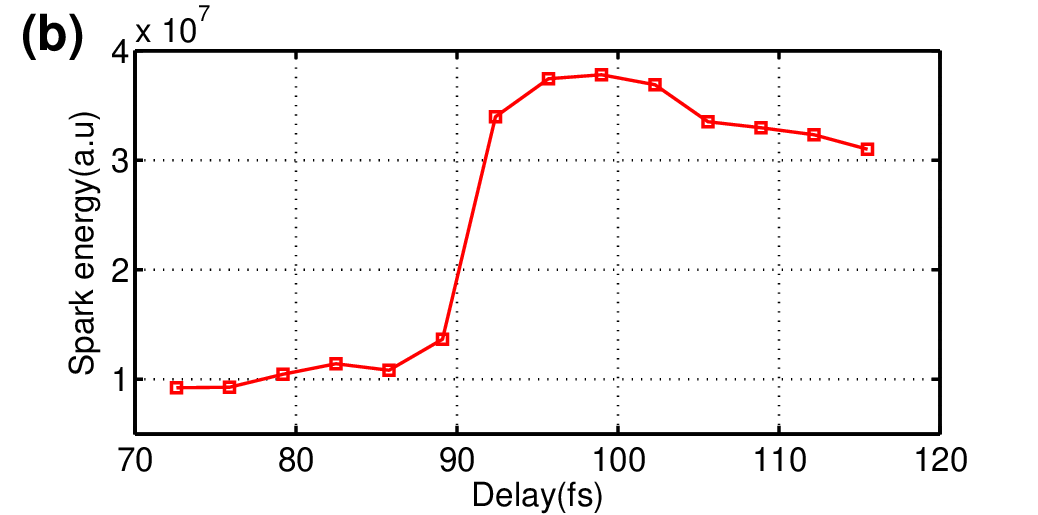}
\caption{(a) Measurement method of flying focus by interfering ionization. (b) Spark energy versus probe delay.
\label{fig:measurement}}
\end{figure}

The adjustment of focus velocity was achieved through manipulation of the compressor grating within the Ti:sapphire CPA laser system, while the measurement process relied on the interfering ionization of the pump and probe. To ensure precise alignment, the spatial and temporal overlap of two laser focuses was meticulously controlled, as evidenced by the heightened intensity of the resulting laser field and the consequent brighter ionizing spark observed in the ambient air. This alignment was iteratively adjusted by monitoring the spark brightness. Specifically, the probe focus was systematically displaced along the trajectory of the flying focus and synchronized with it. Focus velocity was then determined by analyzing the recorded time delays associated with each focus point. Fig.\ref{fig:measurement}(a) provides a visual representation of the methodology. Initially, the probe and flying focuses were aligned at point A. Subsequent displacement of the probe focus by a distance of $\Delta L$ to point B introduced an additional temporal delay of $\Delta t$ to enable interference with the flying focus. Considering that $\Delta t$ encompasses the time taken for the probe to traverse from point B to A, the velocity of the flying focus is given by 

\begin{eqnarray}
\begin{array}{rcl}
v_m=\Delta L/(\Delta t-\Delta L/c),
\end{array}
\label{flyingfocus}
\end{eqnarray}

\begin{eqnarray}
\begin{array}{rcl}
dv_m/v_m=\frac{d(\Delta t)}{\Delta t-\Delta L/c}.
\end{array}
\label{flyingfocus1}
\end{eqnarray}

The precision of the measured focus velocity can be represented by Eq.\ref{flyingfocus1}, wherein it is primarily governed by the synchronization accuracy $d({\Delta t})$ between the pump and probe. To assess $d({\Delta t})$, we employed a CCD to capture ionizing sparks, which exhibit a brightness highly sensitive to the probe delay, as depicted in Fig.\ref{fig:measurement}(b). A decrease of at least $10\%$ in brightness was observed with a probe delay adjustment of 6.6 fs, indicating that the synchronization error is within this temporal range. Considering the full width at half maximum (FWHM) laser duration was approximately 30 fs, the synchronization accuracy attained via interference ionization closely approached 1/5 of the pulse duration. It is noteworthy that $d({\Delta t})$ escalates with increasing pulse duration, thus yielding the minimum $dv_m/v_m$ at the shortest laser duration. As the laser pulse is stretched by the gratings, the measured accuracy of focus velocity diminishes.

\section{Experimental Results}
\begin{figure}[htpb]
\centering
\includegraphics[width=3in]{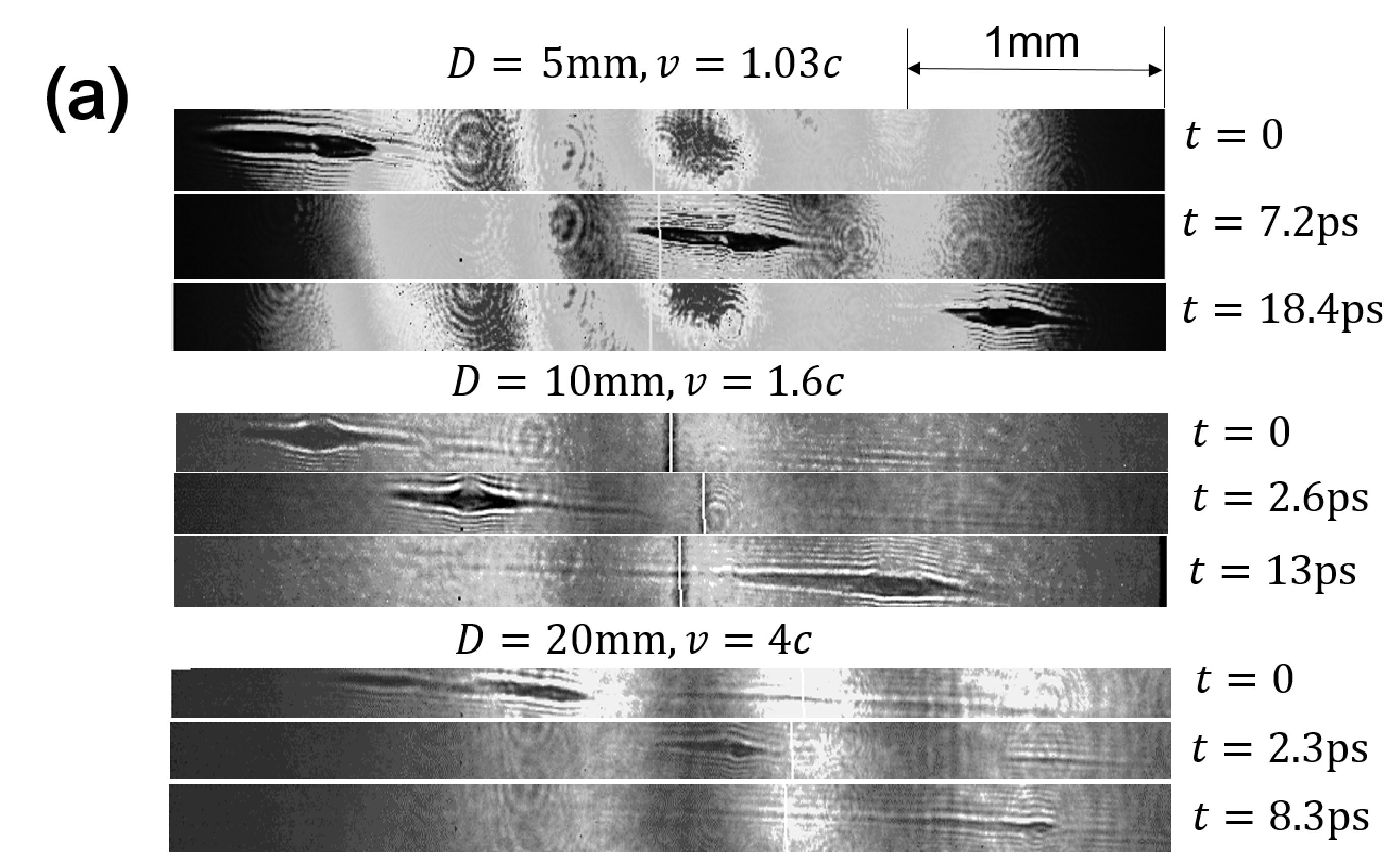}\\
\includegraphics[width=3in]{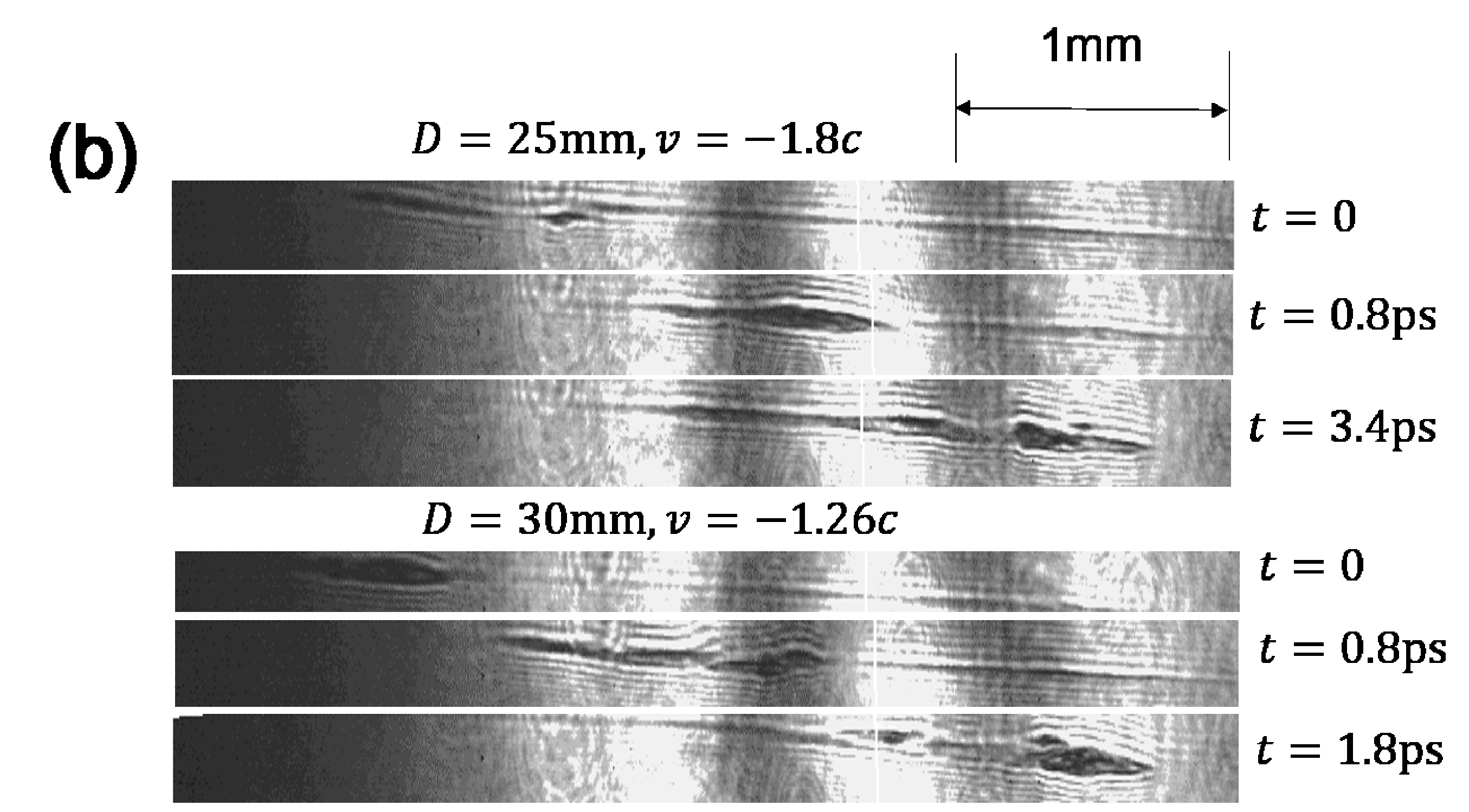}
\caption {(a) Interfering ionized plasma of the probe with positive flying focus. (b) Interfering ionized plasma of the probe with negative flying focus.
\label{fig:flying}}
\end{figure}

The velocity of the flying focus can be adjusted by the pulse chirp, which was given by\cite{Sainte}
\begin{eqnarray}
\begin{array}{rcl}
v=\frac{c}{1-\omega^2_0\beta/\alpha2f^2},
\end{array}
\label{flyingfocus2}
\end{eqnarray}
where $v$ represents the focus velocity, $\omega_0$ denotes the laser center frequency, $\beta$ stands for the pulse chirp, $\alpha$ signifies the group delay of the pulse, and $f$ denotes the focal length of the lens. As outlined in Eq.~\ref{flyingfocus2}, with increasing grating distance in the laser compressor, the pulse chirp $\beta$ increases, causing the focus velocity to first rise from zero to positive superluminal values, and then decrease through negative superluminal values back toward zero.

To compare the experimental results with theoretical predictions, we tested the measurement methodology using pump pulses with varying grating distances. Initially, the pump pulse was set to its shortest duration at $\beta = 0$, corresponding to a focus velocity of $c$ as predicted by Eq.\ref{flyingfocus2}. As shown in Fig.\ref{fig:flying}(a), the measured focus velocity at a grating distance $L = 3$ mm and probe delay $\Delta t = 18.4$ ps was 1.03$c$, demonstrating good agreement with theory. The measurement accuracy, calculated to be approximately $3.6 \times 10^{-4}$ from Eq.~\ref{flyingfocus1}, confirmed the high precision of the method. Next, by increasing the grating distance from 5 mm to 20 mm, the probe delay decreased from 18.4 ps to 8.3 ps, resulting in a corresponding increase in focus velocity from $c$ to approximately 4$c$. Further increases in grating distance produced negative superluminal focus velocities, with -1.8$c$ measured at 25 mm and -1.26$c$ at around 30 mm. However, as the pulse duration extended to approximately 29 ps in these cases, the reduced temporal resolution led to decreased measurement accuracy, making it difficult to resolve a focus velocity of exactly $-c$. Subsequent refinement of focus velocity was performed based on the amplified energy, particularly in the context of its application to stimulated Brillouin scattering (SBS) amplification.

A 3-mm-long plasma channel was generated using the flying focus technique, as shown in Fig.~\ref{fig:flying}(a). Traditionally, producing plasma channels longer than 1 mm has been challenging due to diffractive effects in plasma when using conventional spherical lenses. To overcome this limitation, previous studies have employed line-focus generation techniques based on axicons and diffractive optics \cite{Pai08PRL,Turnbull12,Wu20}. In contrast, the flying focus method offers a simpler and more flexible approach for generating extended plasma channels. By utilizing ZF7-glass spherical lenses—readily available and exhibiting significant chromatic aberration—and combining lenses with different focal lengths, the plasma channel length can be easily tailored.

\begin{figure}[htpb]
\centering
\includegraphics[width=3in]{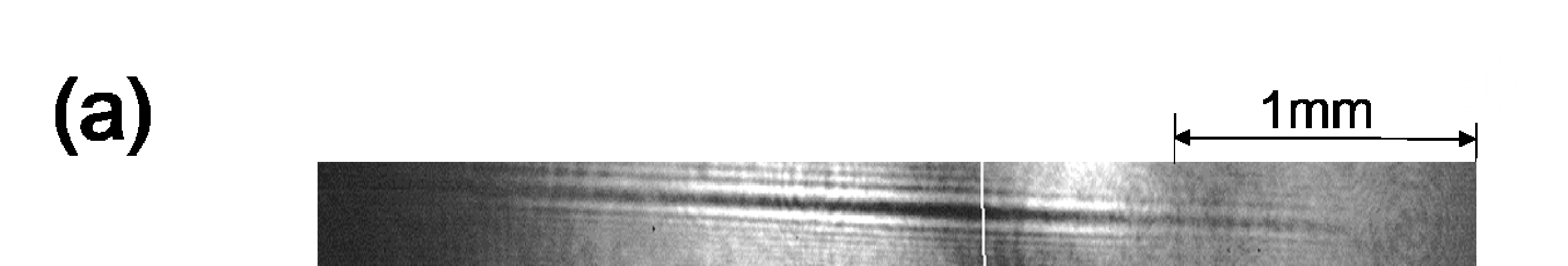}\\
\includegraphics[width=3in]{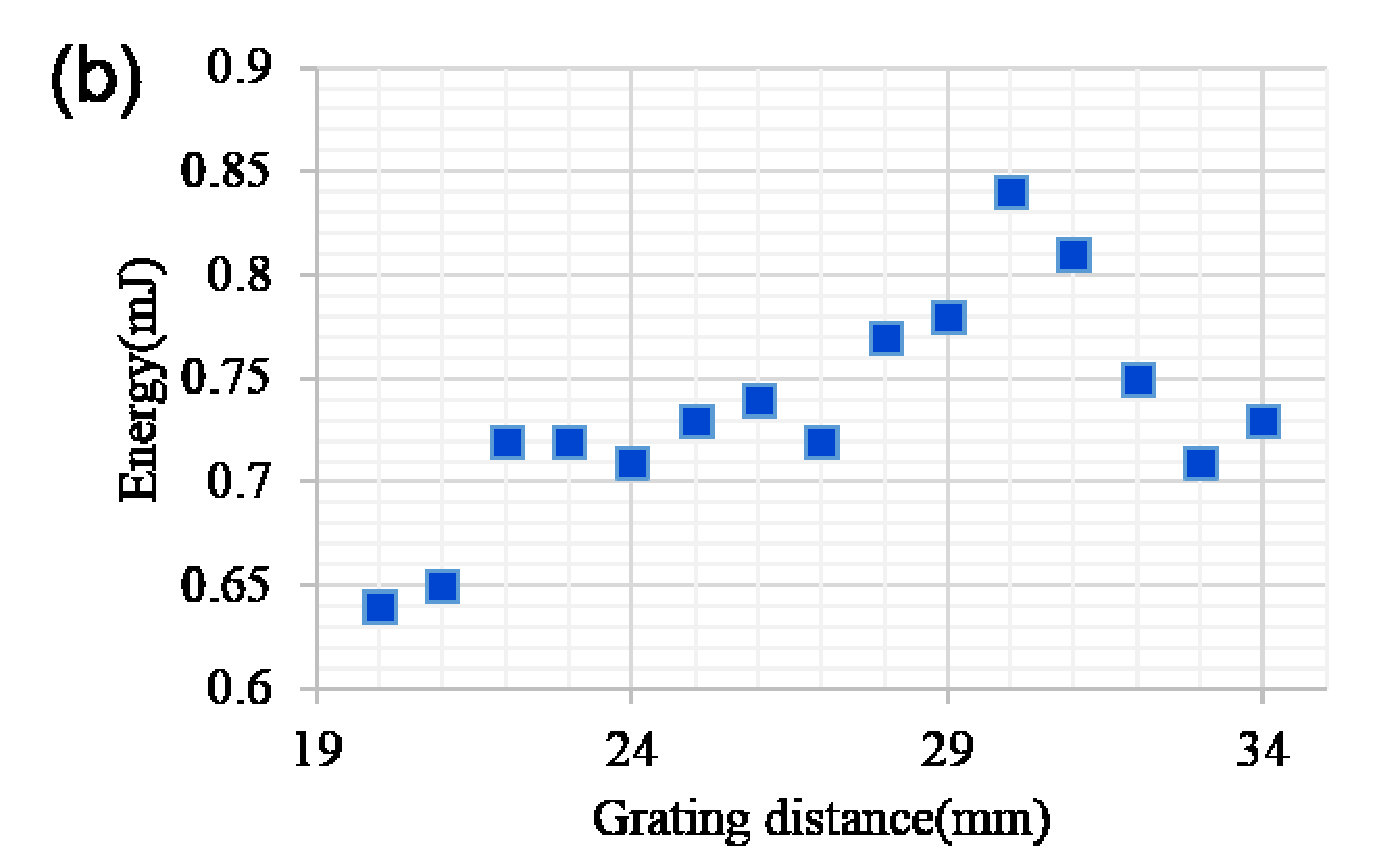}\\
\includegraphics[width=3in]{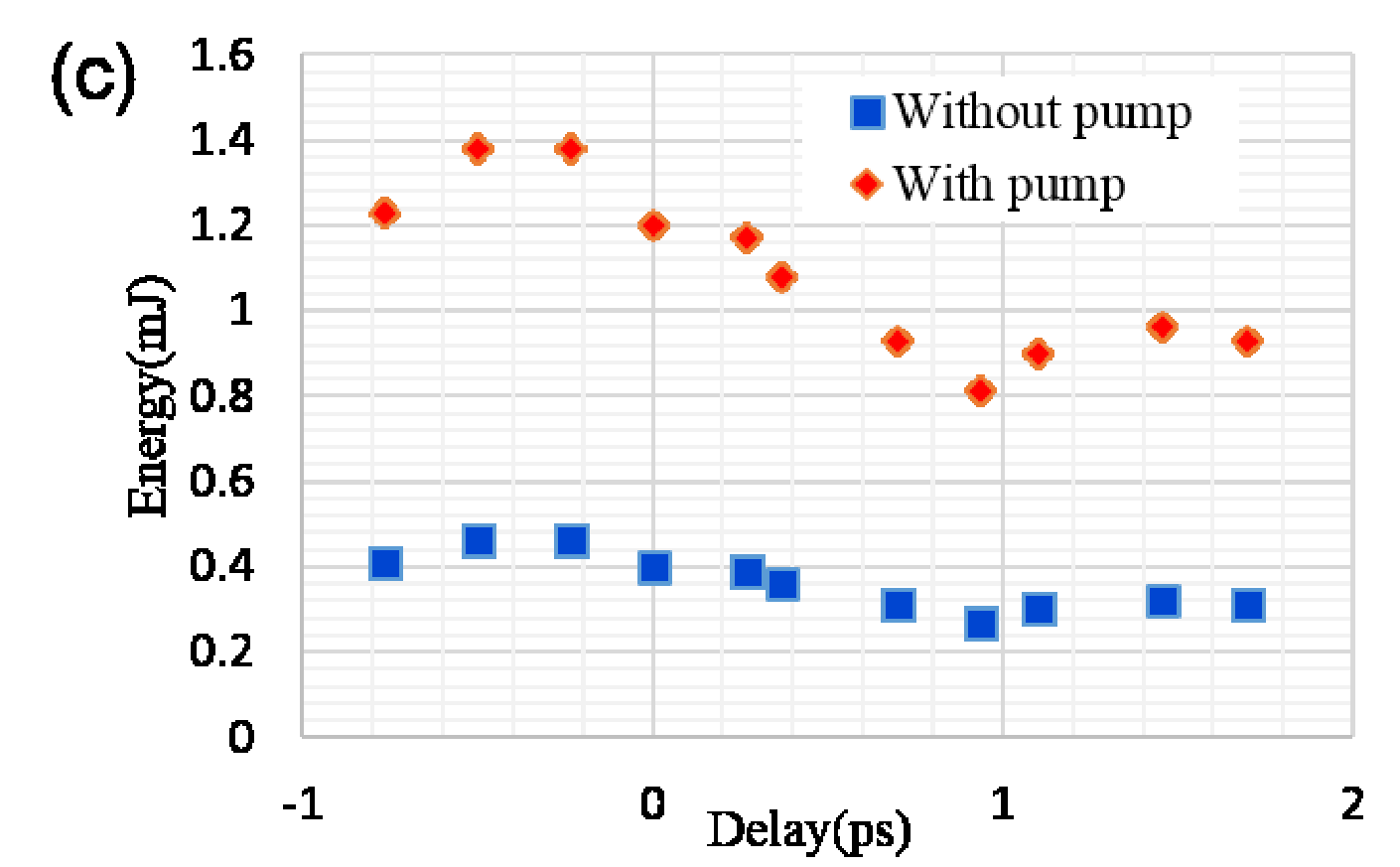}\\
\includegraphics[width=3in]{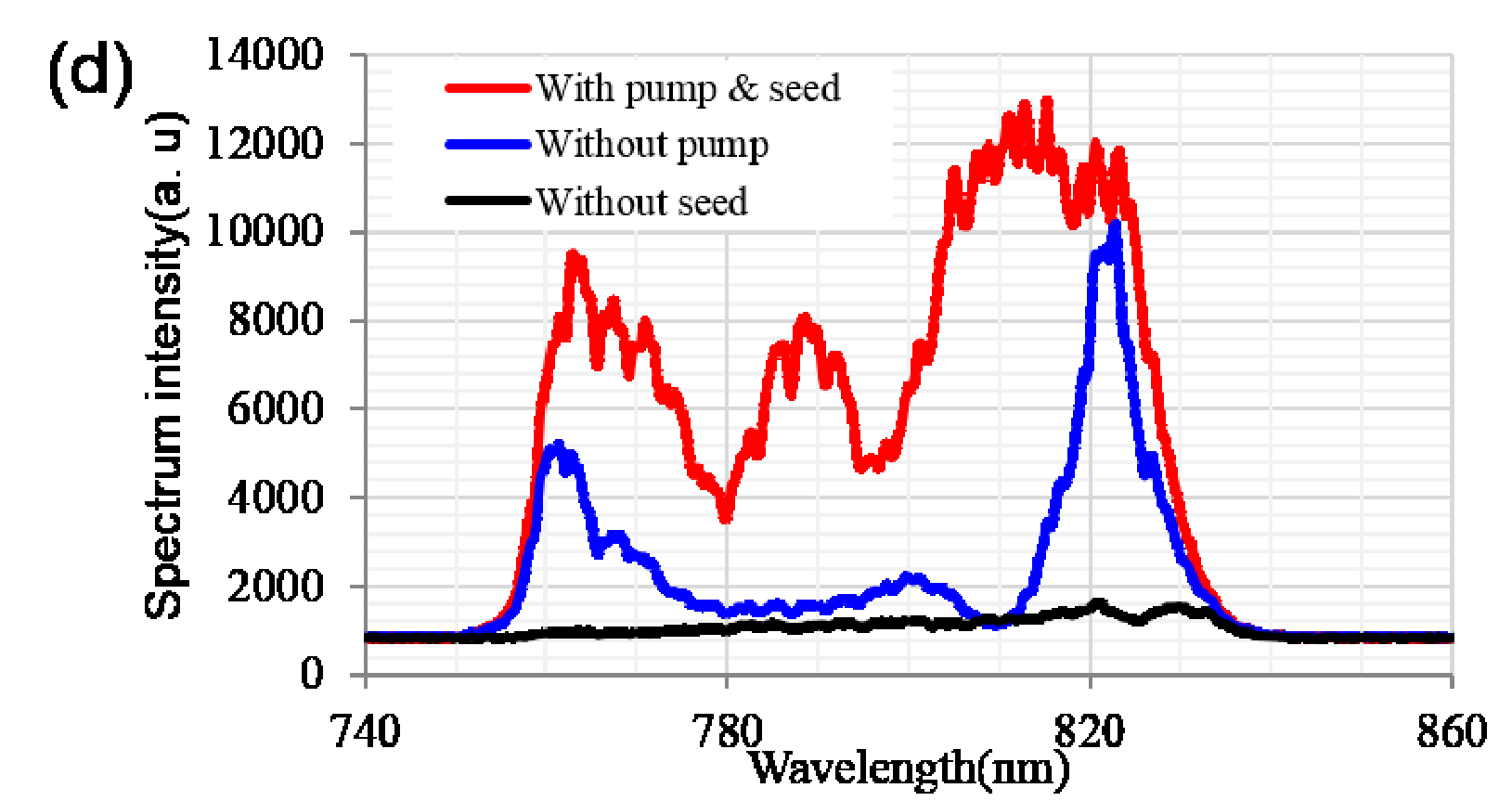}
\caption{(a) Plasma channel created by flying focus. (b) Amplified seed energy at different grating distances.
(c) Amplified seed energy at different seed delays. (d)Output spectra: pump only, seed only, and pump+seed
\label{fig:flying}}
\end{figure}

Following the attainment of a flying focus velocity nearing $-c$, the experiment transitioned to SBS amplification utilizing the identical setup depicted in Fig.\ref{fig:setup1}. In this configuration, the flying focus was employed for the pump beam, while the probe beam served as the seed. To optimize the seed beam, an aperture was utilized to reduce the beam diameter while concurrently increasing the Rayleigh length. Contrary to conventional practices, the seed beam was not subjected to further compression but maintained an equivalent duration to that of the pump beam. Notably, a longer-duration seed has been shown to yield improved results in plasma amplifiers employing flying focus configurations \cite{wu22pre}.

Previous experiments on scSBS in plasma were conducted at laser intensities exceeding $10^{16}\rm{W/cm^2}$, with an interaction length of approximately $\sim250\mu$m \cite{Lancia16,Marques19}. The linear amplification of the electric field in this process is represented by $E_0 \exp(\gamma L)$, where $E_0$ denotes the initial seed electric field, $\gamma$ signifies the growth rate, which scales proportionally to $I^{1/2}$ for weakly coupling SBS and $I^{1/3}$ for strongly coupling SBS, and $L$ represents the interaction length. By implementing the flying focus technique, the interaction length is extended to 3 mm, suggesting that laser intensity can be reduced by over 100 times while maintaining comparable amplification levels. Consequently, the pump and seed intensities utilized in the experiment were $9\times10^{13}\rm{W/cm^2}$ and $5\times10^{13}\rm{W/cm^2}$, respectively. The specific parameters of the laser pulses are provided in Tab.\ref{tab:parameters}.

\begin{table}[htpb]
\centering
\caption{\label{tab:parameters}Parameters of the laser pulses used in the experiment.}
\begin{tabular}{ccccc}
\hline
 Parameters& Energy & Spot size & Duration & Intensity\\
& (mJ) & ($\mu$m) & (ps) & (W$\rm{/cm^2}$)\\
\hline
pump & 6.9 & 18 & 29 &$9\times10^{13}$ \\
Seed & 0.4 & 6 & 29 &$5\times10^{13}$ \\
\hline
\end{tabular}
\end{table}

The experimental results, presented in Fig.\ref{fig:flying}(b)-(d), illustrate the optimization of amplified energy under various grating distances and seed delays. Initially, as the grating distance was adjusted from 20 mm to 34 mm, the amplified seed energy exhibited a peak value of 0.85 mJ at a distance of 30 mm, as depicted in Fig.\ref{fig:flying}(b). Subsequently, the maximum amplified energy of 1.4 mJ was achieved by further optimizing the seed delay to $\sim-0.5$, and then it reduced to approximately 0.4 mJ at the delay of $\sim2$ ps. Therefore, the seed experienced a maximum energy increment of 1 mJ, corresponding to a conversion efficiency of 14.5$\%$, closely aligning with results obtained in the scSBS experiment \cite{Marques19}. It is worth noting that both the pump and seed intensities remained below $10^{14}~\rm{W/cm^2}$, suggesting considerable suppression of plasma instabilities such as filamentation, modulation, and thermal noise. The efficiency of SBS is further underscored by the spectral analysis. As illustrated in Fig.\ref{fig:flying}(d), the seed spectrum underwent significant amplification within the wavelength range of 760 nm to 820 nm. As the amplified seed spectra closely overlapped with that of the pump, it exhibits typical features of SBS.

%\section{Simulation}
%In order compare the experimental result to the theory, we performed the 2D simulation of Brillouin amplification with flying focus. The simulation were carried out in a moving window of $600\lambda \times100\lambda$ for the z-y plane, with cells of $\Delta_z=0.1\lambda$ and $\Delta_y=0.2\lambda$, and $16\times2$ particles for each cell, where $z$ and $y$ are longitudinal and transverse directions, respectively, and $\lambda=1~\mu$m is the pump wavelength. A 3-mm-long and 100-$\mu$m-wide air with 79$\%$ Netron and 21$\%$ oxgen is applied for the background gas. The initial pump focus has a waist of 10 $\mu$m, with the peak amplitude of 0.01. The pulse duration is about 29 ps, with a chirp of. The seed pulse was almost the same as the pump except a small energy. 

\section{Conclusion}

In summary, flying focus was experimentally generated through the accumulation of chromatic aberration in ZF7-glass lenses and adjusted via pulse chirp modulation. Its velocity was quantified using the interfering ionization method, revealing focus velocities ranging from -1.8$c$ to 4$c$, with a measured accuracy of up to $3.6\times10^{-4}$. Notably, the experimental setup, comprising solely lenses and mirrors, offered a remarkably straightforward approach for both creating and measuring flying focus configurations.

By tuning the focus velocity around $-c$, we introduce a novel application of flying focus in SBS within plasma for the first time. While serving as a proof-of-principle experiment, our study showcased the capability of flying focus to initiate SBS in plasma environments, achieving notable results even at laser intensities over 100 times lower than those typically required by conventional static focusing techniques. The resulting amplified seed spectrum exhibited characteristic features indicative of Brillouin scattering, with a maximum conversion efficiency reaching 14.5$\%$.

In the experiment, the aperture of the plasma channel was constrained due to the necessity for a short Rayleigh length in accommodating the flying focus. To address this challenge, a phase plate may be employed to enlarge the focus spot without altering the Rayleigh length of the flying focus. Additionally, the growth rate of the ion-acoustic wave remained low during the experiment, thereby limiting the transfer efficiency. This limitation could potentially be overcome by substituting the uniform gas with a background grating gas. Utilizing a fast-extending plasma grating generated by the flying focus for laser compression, as suggested by recent studies \cite{Zhaohui22,Zhaoli22,wu24prr}, may offer a solution. Such a plasma grating, devoid of a growth process of the plasma wave, has the potential to significantly enhance transfer efficiency.

\section*{Acknowledgments}
This work was partly supported by the National Key Research and Development Program of China (2022YFB3606305), and the National Natural Science Foundation of China (Grant No. 12475248).
\section*{Conflict of interest}
The authors declare no conflicts of interest.
\section*{DATA AVAILABILITY STATEMENT}
Data openly available in a public repository.

\bibliography{ref}% Produces the bibliography via BibTeX.

\end{document}